\UseRawInputEncoding
\documentclass[aps,prb,twocolumn,superscriptaddress]{revtex4-1}
\usepackage[pdftex]{graphicx}
\usepackage{color,soul}
\usepackage{booktabs}
\usepackage{enumitem}
\usepackage{gensymb}
\usepackage{epstopdf}
\usepackage{bm}
\usepackage{amsmath,amsthm,amssymb}
\usepackage{dcolumn}

\begin{document}
\title{Hydrogenation kinetics of metal hydride catalytic layers}

\author{L.J. Bannenberg}
\affiliation{Faculty of Applied Sciences, Delft University of Technology, Mekelweg 15, 2629 JB Delft, The Netherlands}
\email{l.j.bannenberg@tudelft.nl}
\author{H. Schreuders}
\affiliation{Faculty of Applied Sciences, Delft University of Technology, Mekelweg 15, 2629 JB Delft, The Netherlands}

\date{\today}

\begin{abstract}
Catalyzing capping layers on top of metal hydrides are often employed to enhance the hydrogenation kinetics of metal hydride based systems such as hydrogen sensors. Here, we experimentally study the hydrogenation kinetics of capping layers composed of several alloys of Pd and Au as well as Pt, Ni and Ru, all with and without an additional PTFE protection layer using a novel method and under the same set of experimental conditions. Our results demonstrate that doping Pd with Au results in significantly faster hydrogenation kinetics, with response times up to five times shorter than Pd through enhanced diffusion and a reduction of the activation energy. The kinetics of non-Pd based materials turns out to be significantly slower and mainly limited by the diffusion through the capping layer itself. Surprisingly, the additional PTFE layer was only found to improve the kinetics of Pd-based capping materials and has no significant effect on the kinetics of Pt, Ni and Ru. Taken together, the experimental results aid in rationally choosing a suitable capping material for the application of metal hydrides and other materials in a green economy. In addition, the developed method can be used to simultaneously study the hydrogenation kinetics and determine diffusion constants in thin film materials for a wide set of experimental conditions.
\end{abstract}
\maketitle


\section{Introduction}
Hydrogen has the potential to become an important energy vector in a future green economy \cite{brandon2017,natmat2018,abe2019,glenk2019}. Metal hydrides have been projected to play an important role in such a hydrogen-powered economy. While these materials have traditionally been studied as materials for hydrogen storage \cite{rusman2016,schneemann2018,bannenberg2020review}, other applications as in switchable mirrors \cite{huiberts1996,maiorov2020}, hydrogen purifying membranes \cite{nishimura2002,dolan2018}, fuel cells \cite{lototskyy2017} and especially hydrogen sensors became increasingly prominent over the last years \cite{hubert2011,wadell2014,bannenberg2020,bannenberg2020review,darmadi2020,koo2020,chen2021}. In all of these applications, fast kinetics are extremely important. However, while the diffusion of hydrogen is in most materials relatively fast, which would as such enable high (de)sorption rates even at room temperature, the observed kinetics are generally speaking much slower. Therefore, catalyzing capping on top of the active material are often used. These layers promote the dissociation of hydrogen molecules (H$_2$) into atomic hydrogen at their surface and may also protect the layer from e.g. oxidation and poisonous chemical species. 


Especially for metal hydride based hydrogen sensors, capping layers play an important role to accelerate the kinetics or as additional protection to avoid chemical cross-sensitivity. One of the many different metal hydride based sensors are optical hydrogen sensors with a separate hydrogen dissociation and sensing functionality have been developed \cite{slaman2007,boelsma2017,bannenberg2019,bannenberg2020review}. In such optical hydrogen sensors, the sensing layer absorbs hydrogen when it is exposed to an environment in which hydrogen, while a separate capping layer on top catalyzes the hydrogen dissociation and provides protection against chemical poisoning. In turn, the absorption of hydrogen by the sensing layer changes the optical properties of this layer. By e.g. measuring the reflectivity or transmission of the sensing layer, one can determine the hydrogen pressure or concentration in the environment of the sensor. 

Whereas in fuel cells Pt is typically used to catalyze the hydrogen dissociation \cite{antolini2009,holton2013}, in capping layers for metal hydride based hydrogen sensors and in switchable mirrors Pd is often the material of choice. Pd can readily dissociate hydrogen at room temperature \cite{hubert2011,wadell2014,bannenberg2020,bannenberg2020review,darmadi2020}, and the application of a 30~nm polytetrafluoroethylene (PTFE) or other polymeric layers on top of Pd can significantly increase it's chemical selectivity and enhance the kinetics. These effects may be due to a lower activation energy or to more active sites on the surface remaining available to dissociate the hydrogen \cite{ngene2014,delmelle2016,nugroho2019}. 

However, Pd has some major drawbacks as a catalyzing layer. The first-order transition from the dilute $\alpha$-PdH$_x$ to the higher concentration PdH$_x$ $\beta$-phase occurring upon hydrogen sorption renders the optical response highly hysteric, which is problematic for hydrogen sensors. Most importantly, the high energy barriers associated with the nucleation of domains when the $\beta$-phase is formed results in relatively long response times. Furthermore, long-term stability issues due to the formation of cracks, buckling and delamination as a result of the large volumetric expansion upon hydrogenation have been reported for thin layers of Pd \cite{pivak2009,pivak2011}. 

Capping layers different from Pd are thus required and in this paper we consider two approaches to find these alternative capping layers. The first approach is to alloy Pd with other elements like Au, which suppresses the first-order phase transition for sufficiently large Au concentrations \cite{luo2010,nugroho2019,bannenberg2019PdAu}, increase the hydrogen solubility \cite{luo2010,nugroho2019,bannenberg2019PdAu} and may decrease the activation energy \cite{namba2018}. The second approach is to consider single-element layers made of elements other than Pd as Pt, Ni and Ru, i.e. materials that itself hardly absorb hydrogen but can catalyze the dissociation and recombination reaction of hydrogen. Albeit it is well known that these single element layers may catalyze hydrogen absorption, a dedicated and comprehensive study in which a series of capping layers are tested under exactly the same circumstances and under a wide variety of conditions (temperature, pressure) is at present unavailable. 

The purpose of this paper is to fill this void by experimentally studying the hydrogenation kinetics of various capping layers composed of (i) several alloys of Pd and Au as well as (ii) single element Pt, Ni and Ru, all with and without a PTFE protection layer and under exactly the same experimental conditions. We do this by using a newly developed method based on studying the optical transmission of capped films of Ta. Here, the Ta serves as an indicator layer owing to its large changes of the optical properties and high diffusivity of hydrogen. These changes are not only large but also gradual: the optical transmission continuously decreases when the hydrogen pressure is increased, allowing us to assess the kinetics of the capping layers over a wide pressure (and temperature) range \cite{bannenberg2019,bannenberg2021}. As such, we can in this way systematically measure the hydrogenation kinetics of many different samples simultaneously and under a wide range of experimental conditions.

The results presented in this paper demonstrate that doping Pd with Au significantly fastens the hydrogenation kinetics under all probed experimental conditions, with response times up to five times shorter than Pd. These fastened kinetics result both from enhanced diffusion and a reduction of the activation energy. From the non-Pd based capping layers, only Ni turns out to be competitive at room temperature with response times slightly longer than Pd. Surprisingly, PTFE was only found to improve the kinetics of Pd-based capping materials and had no significant effect on the kinetics of Pt, Ni and Ru. As such, these results aid in rationally selecting a suitable capping material for the application of metal hydrides in a green economy. In a wider perspective, the method developed in this paper can be used to simultaneously study the hydrogenation kinetics and determine diffusion constants in thin film materials for a wide set of experimental conditions.

\begin{figure*}[tb]
\begin{center}
\includegraphics[width= 0.9\textwidth]{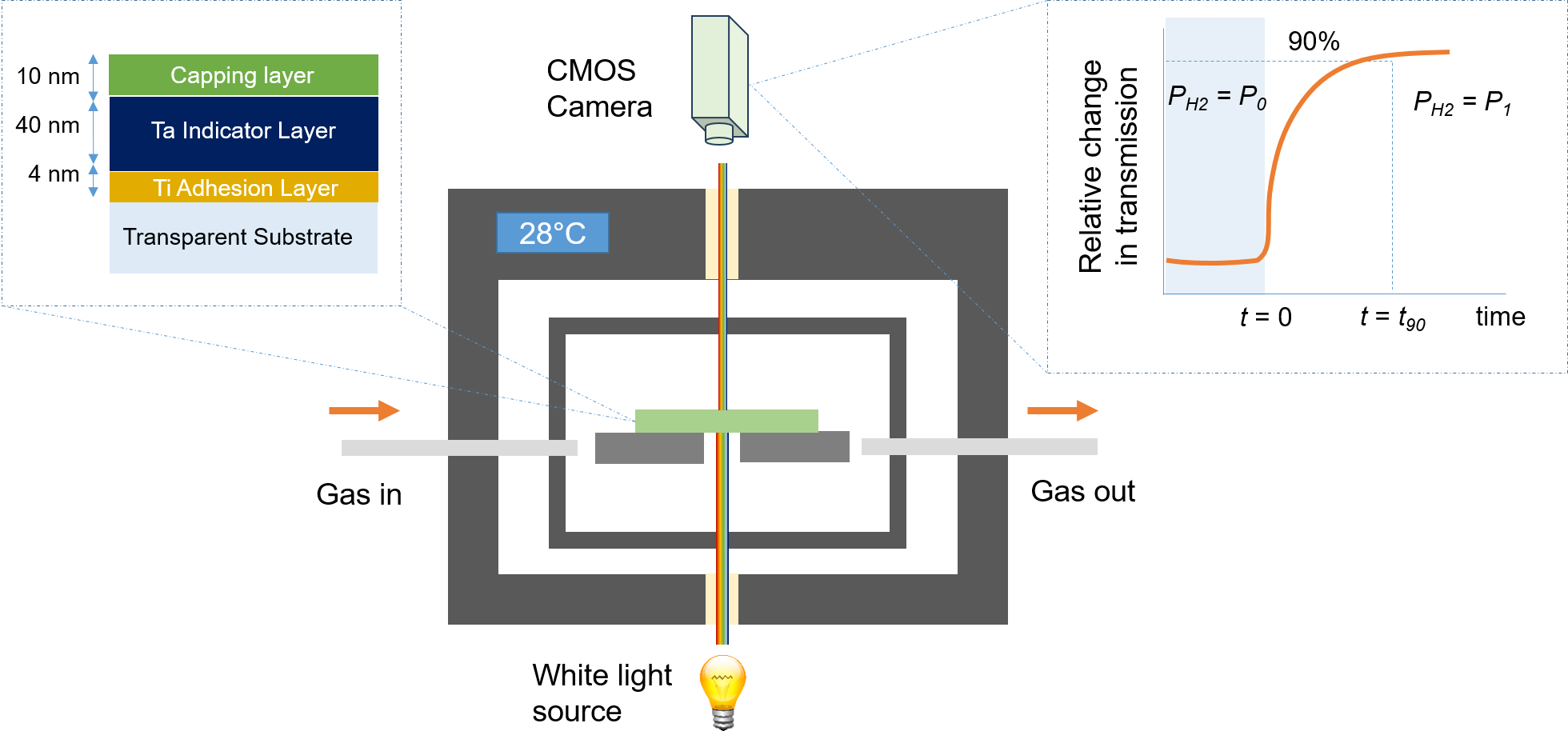}
\caption{Schematic illustration of the method used to study the hydrogenation kinetics of the capping layers. In this method, we use a multilayer thin film consisting of a Ti adhesion layer, a Ta indicator layer and the capping layer of interest (see left inset). The thin film is positioned inside a hydrogenation chamber, itself located in a temperature controlled environment. We monitor the time-dependence of the changes in optical transmission of the sample with a CMOS camera. Upon a change in hydrogen pressure, the optical transmission of the Ta layer changes, which we use to determine the response time $t_{90}$ that is here is defined as the time to reach 90\% of the total signal (see right inset). }
\label{ExperimentalSetup}
\end{center}
\end{figure*}

\section{Experimental}
\subsection{Experimental method and sample composition}
To study the hydrogenation kinetics of the capping layers, we use a multilayer thin film consisting of a Ti adhesion layer, a Ta indicator layer and the capping layer of interest. As schematically illustrated in Fig. \ref{ExperimentalSetup}, we monitor (the changes) in optical transmission of the Ta layer inside a hydrogenation chamber as a function of time. Upon a change in hydrogen pressure, the optical transmission of the Ta layer changes, from which we determine the response time. As previous measurements indicate that the logarithm of the relative change in transmission of Ta is proportional to the hydrogenation of the layer \cite{bannenberg2019}, this allows an absolute determination of the hydrogenation rate.

The reasons for selecting Ta as our indicator layer is threefold. First, thin film Ta has a large hydrogen solubility window and gradual hydrogenation over a large partial pressure range without any hysteresis and within one single thermodynamic phase \cite{bannenberg2021}. As such, it allows us to examine the (de)hydrogenation kinetics of capping layers over both a large temperature and pressure window without the presence of any sluggish, hysteretic or irreversible phase transition. Second, the changes in optical transmission of Ta upon exposure to hydrogen are large and proportional to the hydrogenation of the layer over the entire solubility range \cite{bannenberg2019}, allowing for high-resolution response-time measurements and absolute calibration of the hydrogenation rate. Third, previous measurements indicate that diffusion of hydrogen through body centered cubic (bcc) Ta thin films is, as for bulk \cite{fukai2006}, fast \cite{bannenberg2019}, occurring at the sub-second timescale, and will thus not influence the total response times.

We measure the changes in optical transmission in the same way as in hydrogenography \cite{gremaud2007hydrogenography} using an Imaging Source 1/2.5" Aptina CMOS 2592 $\times$ 1944 pixel monochrome camera equipped with an Edmunds Optics 55-906 lens, i.e. the same camera as used in ref. \cite{bannenberg2016}, which has an maximum acquisition frequency of 20~Hz. Five Philips MR16 MASTER LEDs (10/50~W) with a color temperature of 4,000~K are used as a light source and provide a white spectrum (Fig. \ref{SpectrumLED}). The transmission is averaged over an area of 180 $\times$ 180 pixels, corresponding to about 80~mm$^2$. A reference sample is used to compensate for fluctuations of the LED white light source. The partial hydrogen pressures of 10$^{-1}$ $<$ $P_{H2}$ $<$ 10$^{+6}$~Pa are obtained by using 0.1\%, 3.5\% and 100\% H$_2$ in Ar gas mixtures. During the experiments a minimum gas flow of 20~s.c.c.m. is always maintained.

The response time $t_{90}$ is defined as the time to reach 90\% of the total signal. To measure the hydrogen absorption response times, the pressure was increased from the base pressure of 0.9~mbar \textit{total} pressure to the pressure of interest (maximum 100~mbar total pressure). As a result of the large partial hydrogen pressure range investigated, three different hydrogen base pressures were used for the three H$_2$ in Ar gas mixtures. The pressure was stepwise increased from $P_{H2}$ = 0.090~Pa for response times measured for $P_{H2}$ $\leq$ 10~Pa, from $P_{H2}$ = 5~Pa for response times measured for 9~Pa $\leq$ $P_{H2}$ $\leq$ 350~Pa, and from $P_{H2}$ = 90~Pa (the regions partly overlap) for response times measured for $P_{H2}$ $>$ 150~Pa. The minimum time required to achieve a stable pressure, i.e. the shortest response time that can be reliably detected, was found to be 0.3~s. 

\subsection{Sample preparation and characterization}
All samples are produced with magnetron sputtering and composed of a 4~nm Ti adhesion layer, a 40~nm Ta indicator layer that generates (the fast majority of) the optical contrast and a 10~nm capping layer and an optional 30~nm PTFE layer. As a capping layer, we consider Pd$_{1-y}$Au$_y$ layers with $y$ = 0, 0.1, 0.2, 0.3, 0.4, 0.5, Ni, Pt and Ru. Direct current (DC) magnetron sputtering of the metallic layers is performed in 0.3~Pa of Argon and inside an ultrahigh vacuum chamber (AJA Int.) with a base pressure of 10$^{-10}$~Pa. The layers were deposited on 10 $\times$ 10~mm$^2$ quartz substrates (thickness of 0.5~mm and surface roughness $<$ 0.4~nm) which were rotated to enhance the homogeneity of the deposited layers. Typical deposition rates include 0.11-0.22~nm s$^{-1}$ (22-50~W DC) for Pd, 0.10~nm s$^{-1}$ (125~W DC) for Ta, 0.05~nm s$^{-1}$ (100~W DC) for Ti, 0.11-0.22~nm~s$^{-1}$ (6-25~W DC) for Au and 0.07~nm s$^{-1}$ (100~W DC) for Ni, 0.12~nm s$^{-1}$ (50~W DC) for Pt and 0.13~nm s$^{-1}$ (100~W DC) for Ru. These rates were determined by sputtering each target independently at a fixed power over a well-defined time interval. Subsequently, X-ray reflectometry (XRR) was used to estimate the layer thickness of this reference sample (see below for experimental details), from which the sputter rate was computed. Different from the other layers, PTFE was deposited by radiofrequency magnetron sputtering in 0.5~Pa of Ar. The thickness of the reference film was in this case measured with a DekTak3 profilometer. While all targets were pre-sputtered for at least 5~min, the Ta target was pre-sputtered for at least 240~min to avoid possible contamination from the tantalum oxide and nitride layers possibly present at the surface of the target.

The structure and thickness of all samples was verified with X-ray diffraction (XRD) and XRR (Fig. \ref{XRR} and Fig. \ref{XRD}) using a Bruker D8 Discover (Cu-K$\alpha$ $\lambda$ = 0.1542~nm). The XRR measurements were fitted with GenX3 \citep{bjorck2007} to obtain estimates for the layer thickness, roughness and density of the thin films. These fitted parameters are reported in Table \ref{XRRfits} and reveal that the deviation of the layer thickness between the different samples is at maximum 3\%, that the density of the various layers are consistent with the literature value reported for bulk materials, and that the roughness of the surface of the capping layer is at maximum 1.5~nm. XRD measurements reveal that all capping layers crystallize in an face-centered cubic structure and are textured with the $\langle$111$\rangle$ - crystallographic direction in the out-of-plane direction. The lattice constants for the Pd$_{1-y}$Au$_y$ samples follow Vergard's law. 

Before the measurements, we exposed the thin films to three cycles of hydrogen with a maximum pressure of $P_{H2}$  = 10$^{+6}$~Pa at $T$ = 28~$\degree$C. Reproducible results in terms of both hydrogenation kinetics and the absolute hydrogen solubility are obtained from the second cycle onwards. The difference between the first and subsequent cycles are common to thin film metal hydrides: Generally speaking, a few cycles of exposure to hydrogen are required to show reproducible results due to a settling of the microstructure (see also Fig. \ref{XRD}). In particular, we note that during the first exposure to hydrogen, the Ni capped sample only hydrogenated after 12~h of exposure to $P_{H2}$ = 10$^{+6}$~Pa at room temperature, while much faster responses were observed for subsequent cycles. This suggests a substantial rearrangement of the microstructure upon first exposure to hydrogen.

\begin{figure*}[tb]
\begin{center}
\includegraphics[width= 0.75\textwidth]{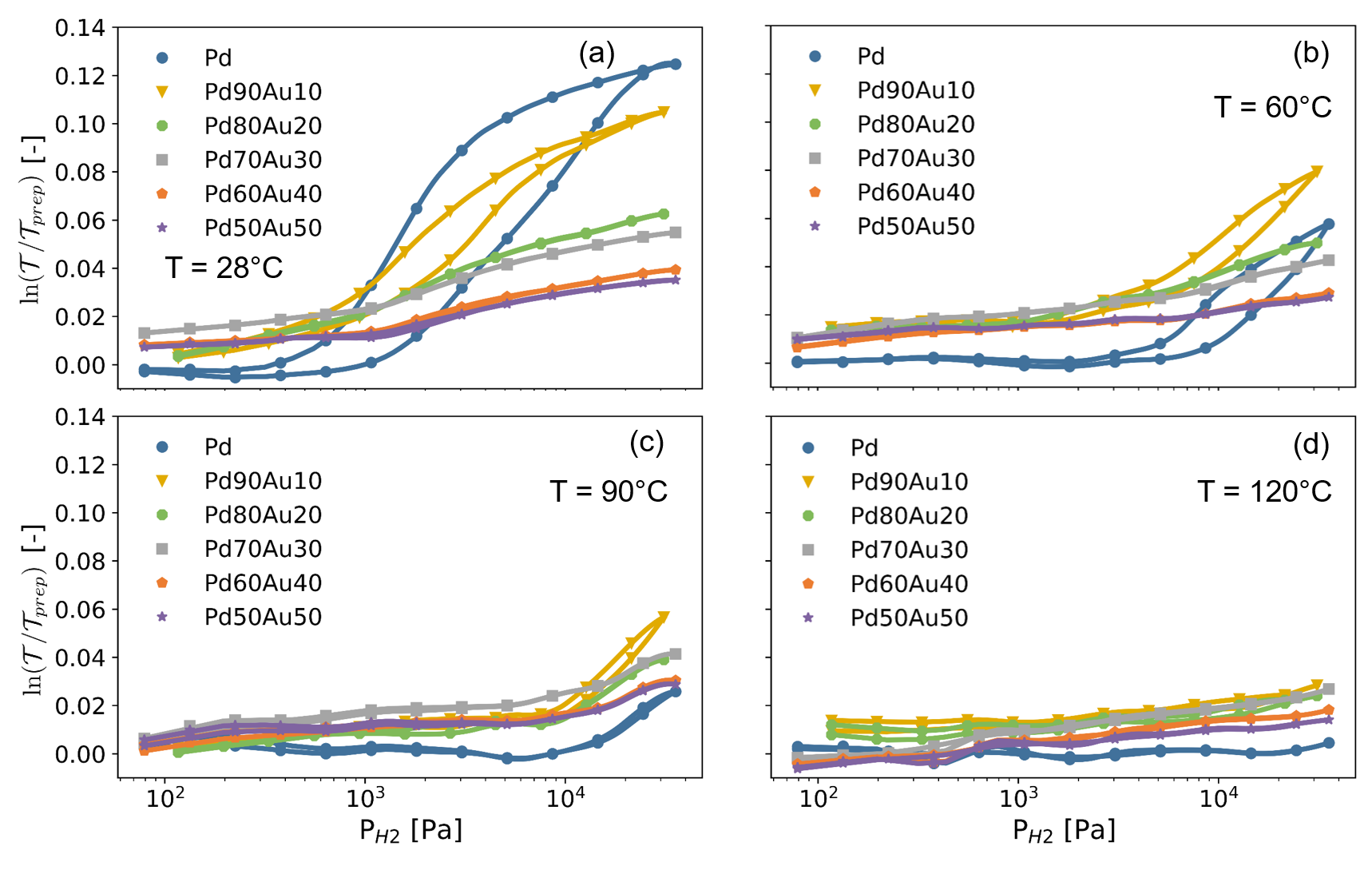}
\caption{Partial hydrogen pressure dependence of the optical responses of 10~nm Pd$_{1-y}$Au$_{y}$ capping layers at (a) $T$ = 28, (b) 60, (c) 90 and (d) 120 $\degree$C. The response was measured by stepwise increasing and decreasing the pressure between $P_{H2}$ = 1 $\times$ 10$^{1}$ and 4 $\times$ 10$^5$~Pa. The optical response is measured as the natural logarithm of the changes of the white light optical transmission $\mathcal{T}$ relative to the transmission of the as-prepared film ($\mathcal{T}_{prep}$) and has been computed by subtracting the response of the 4~nm Ti adhesion layer and 40~nm Ta layer as reported in ref. \cite{bannenberg2021} from the Pd$_{1-y}$Au$_{y}$ capped 40~nm Ta thin films with a 4~nm Ti adhesion layer. As the changes in transmission of the capping layer are relatively small w.r.t. of the Ta indicator layer and the measurement time was relatively long, this caused some long-term fluctuations in the determined changes in transmission of the capping layers. Previous neutron reflectometry measurements indicate that $\ln$($\mathcal{T}/\mathcal{T}_{prep})$ is proportional to the the hydrogen content $x$ in the Pd$_{1-y}$Au$_{y}$H$_x$ layer \cite{bannenberg2019PdAu}. }
\label{OpticalResponse}
\end{center}
\end{figure*}
\begin{figure*}[tb]
\begin{center}
\includegraphics[width= 1\textwidth]{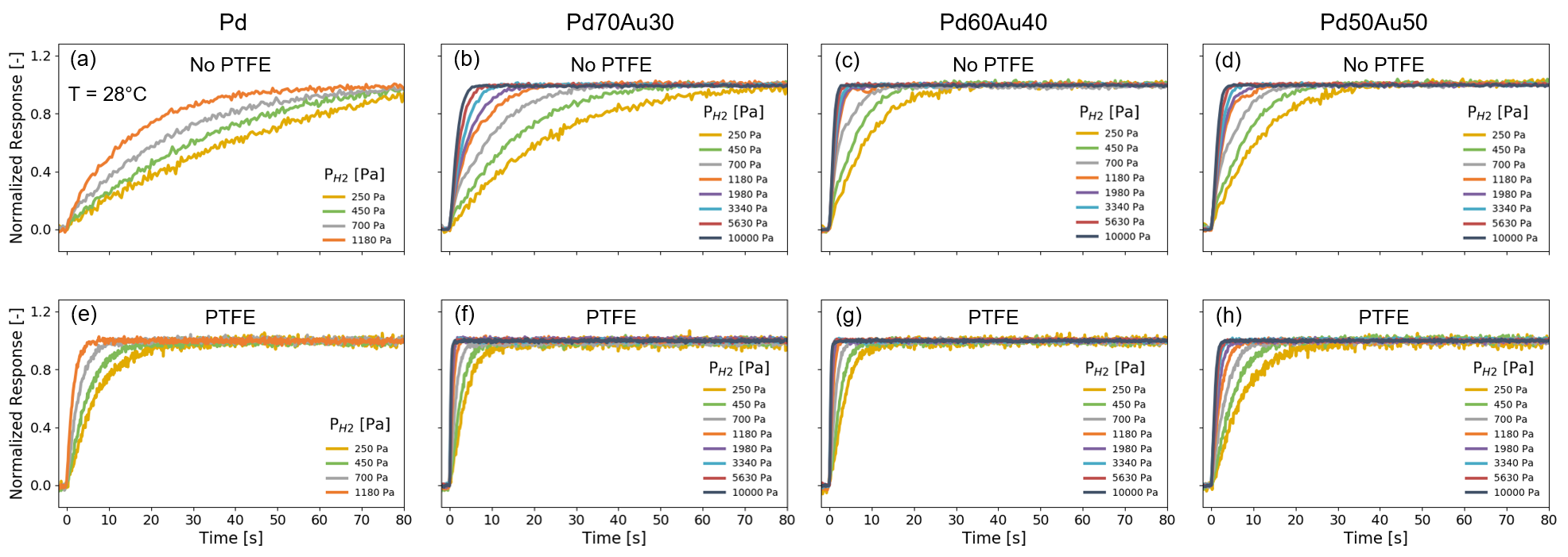}
\caption{Normalized optical transmissions showing the absorption kinetics of a 40~nm Ta thin film with a 4~nm Ti adhesion layer capped with a 10~nm (a,e) Pd, (b,f) Pd$_{0.7}$Au$_{0.3}$, (c,g) Pd$_{0.6}$Au$_{0.4}$ and (d,h) Pd$_{0.5}$Au$_{0.5}$ layer to a series of pressure steps between $P_{H2}$ = 0.5 10$^2$~Pa and the partial hydrogen pressure indicated. The samples in (e-h) are covered with a 30~nm PTFE layer. }
\label{RawIncreasingRT}
\end{center}
\end{figure*}

\begin{figure*}[tb]
\begin{center}
\includegraphics[width= 1\textwidth]{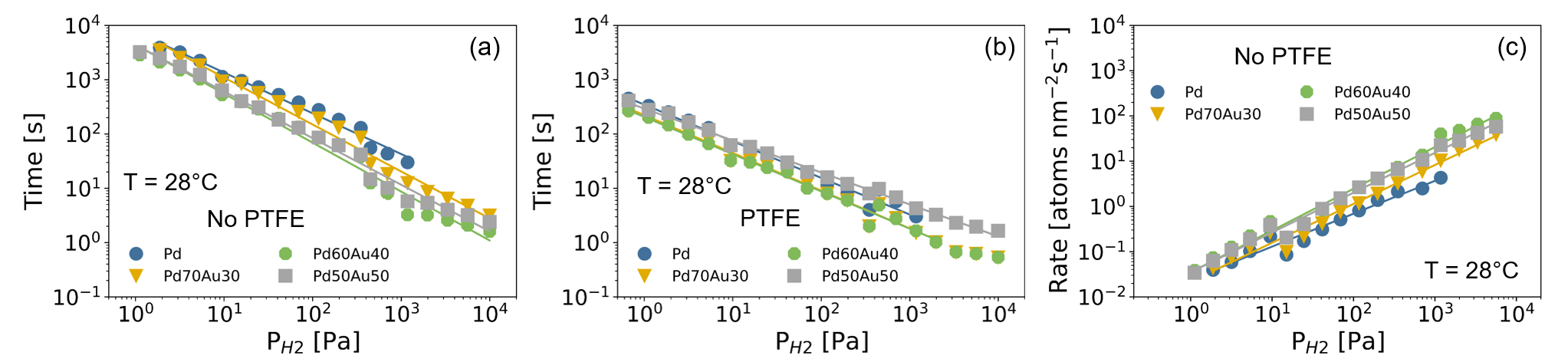}
\caption{Hydrogen absorption kinetics of 40~nm Ta thin films capped with 10~nm Pd$_{1-y}$Au$_{y}$ at $T$ = 28~$\degree$C. Hydrogen pressure dependence of the response time (a) without and (b) with a 30~nm PTFE layer on top of the capping layer. The solid lines indicate fits to a power law ($t_{response} = a\cdot P_{H2}^\gamma$). The response time is defined as the time to reach 90\% of the total signal in Fig. \ref{RawIncreasingRT}. (c) Pressure dependence of the hydrogen absorption rate $R$ computed according to eq. \ref{eq1}. The solid lines indicate fits to a power law ($R = a\cdot P_{H2}^\gamma$) for which the exponents $\gamma$ are reported in Table \ref{exponentsincreasing}.}
\label{PressureDependenceIncreasingRT}
\end{center}
\end{figure*}

\begin{figure*}[tb]
\begin{center}
\includegraphics[width= 1\textwidth]{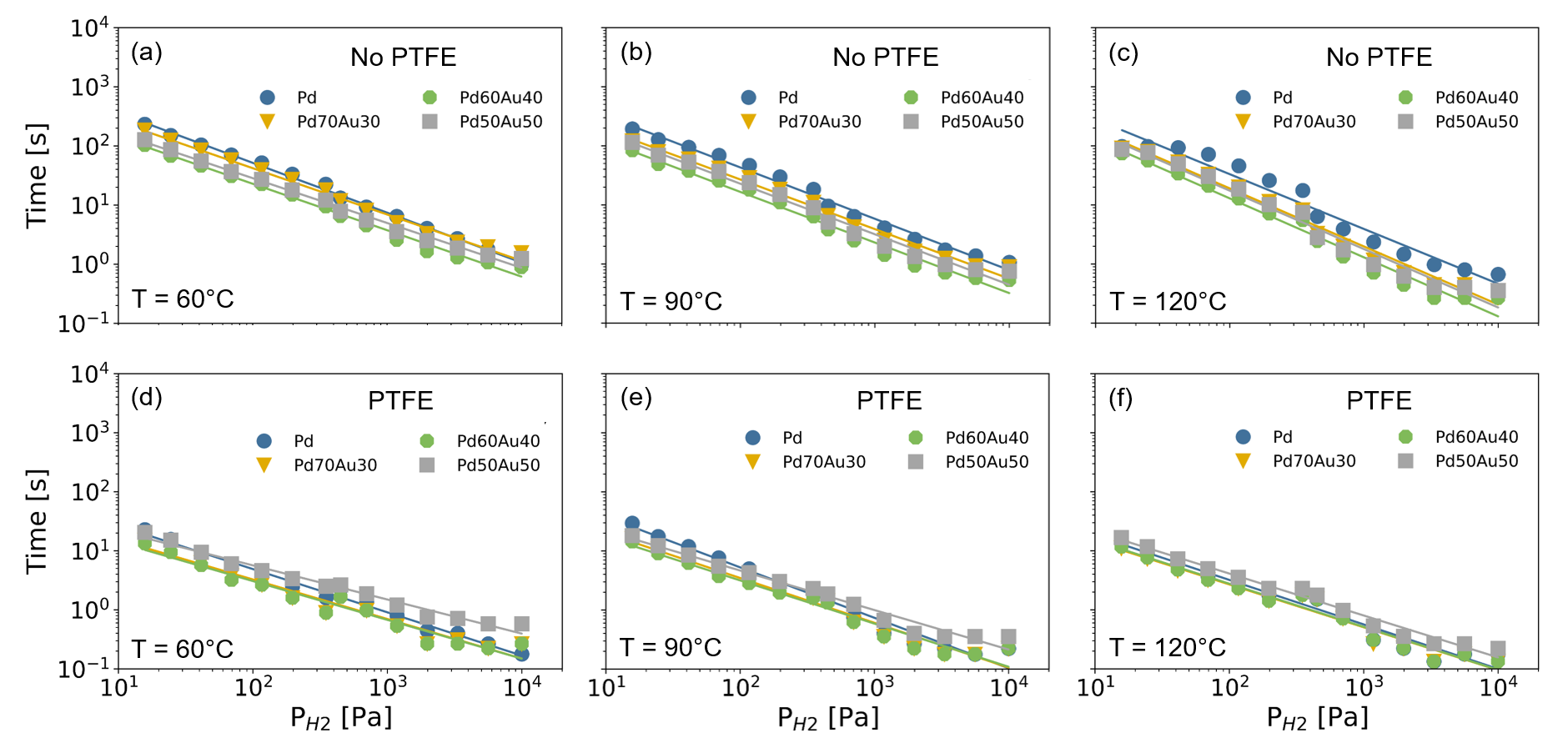}
\caption{Hydrogen pressure dependence of the hydrogen absorption response times of 40~nm Ta thin films capped with 10~nm Pd$_{1-y}$Au$_{y}$ at (a,d) $T$ = 60, (b,e) 90 and (c,f) 120~$\degree$C (a-c) without and (d-f) with a 30~nm PTFE layer on top of the capping layer. The solid lines indicate fits to a power law ($t_{response} = a\cdot P_{H2}^\gamma$). The response time is defined as the time to reach 90\% of the total signal in Fig. \ref{RawIncreasingRT}. }
\label{PressureDependenceIncreasingHighT}
\end{center}
\end{figure*}

\begin{figure*}[tb]
\begin{center}
\includegraphics[width= 1\textwidth]{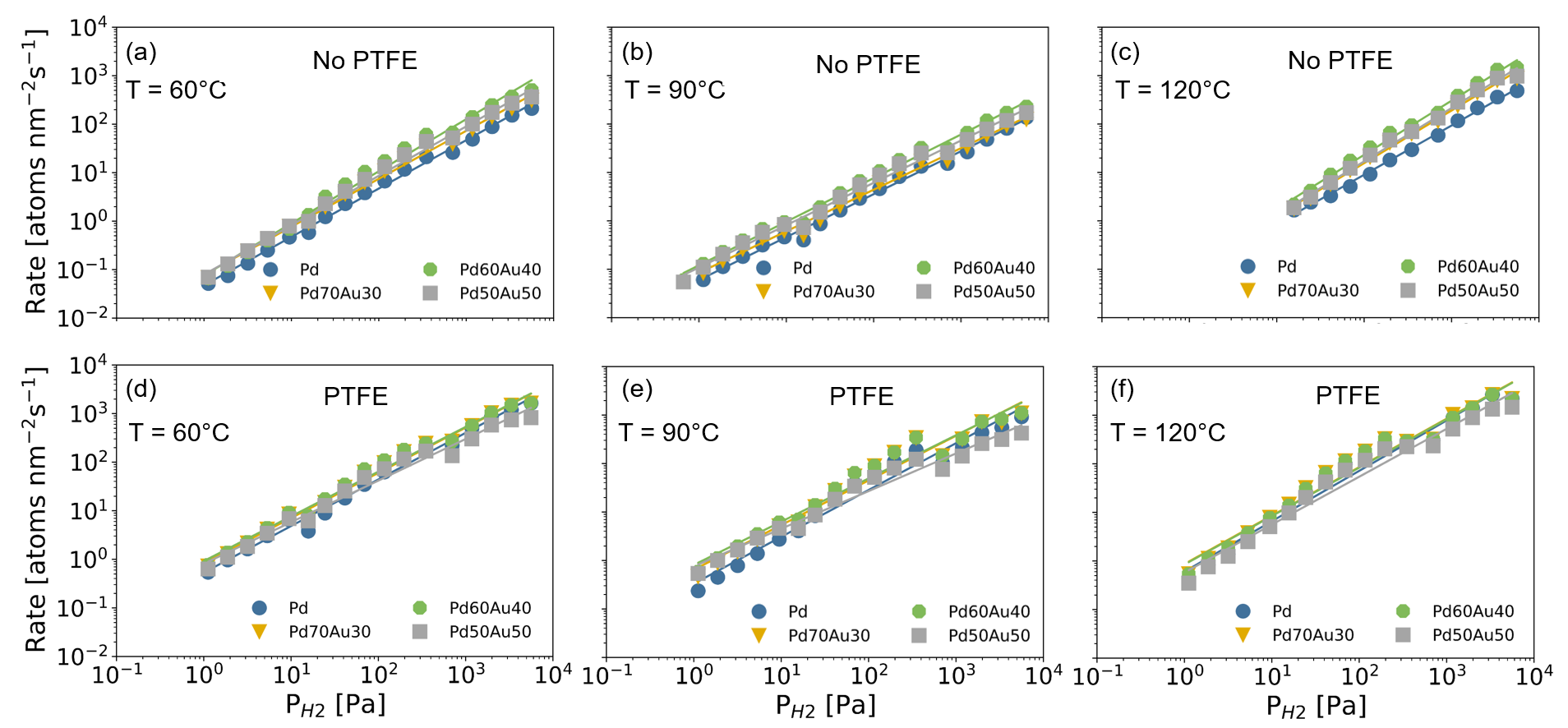}
\caption{Hydrogen pressure dependence of the hydrogen absorption rate of 40~nm Ta thin films capped with 10~nm Pd$_{1-y}$Au$_{y}$ at (a,d) $T$ = 60, (b,e) 90 and (c,f) 120~$\degree$C (a-c) without and (d-f) with a 30~nm PTFE layer on top of the capping layer. The hydrogen absorption rate $R$ is computed according to eq. \ref{eq1}. The solid lines indicate fits to a power law ($R = a\cdot P_{H2}^\gamma$) for which the exponents $\gamma$ are reported in Table \ref{exponentsincreasing}. }
\label{PressureDependenceRateIncreasingHighT}
\end{center}
\end{figure*}

\begin{figure*}[tb]
\begin{center}
\includegraphics[width= 1\textwidth]{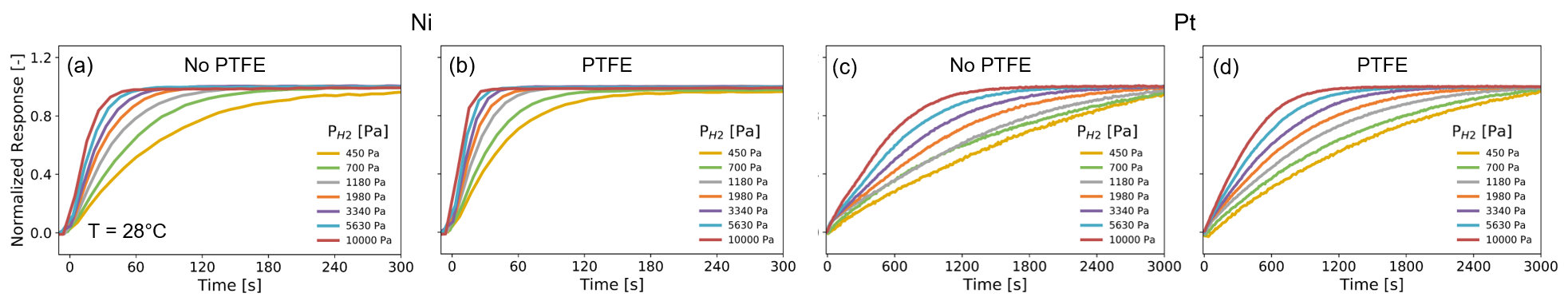}
\caption{Normalized optical transmissions showing the absorption kinetics of a 40~nm Ta thin film with a 4~nm Ti adhesion layer capped with a 10~nm Ni (a,b) and Pt, (c,d) layer to a series of pressure steps between $P_{H2}$ = 0.5 10$^2$~Pa and the partial hydrogen pressure indicated. The samples in (b,d) are covered with a 30~nm PTFE layer. }
\label{RawIncreasingRT_Ni_Pt}
\end{center}
\end{figure*}

\begin{figure*}[tb]
\begin{center}
\includegraphics[width= 0.9\textwidth]{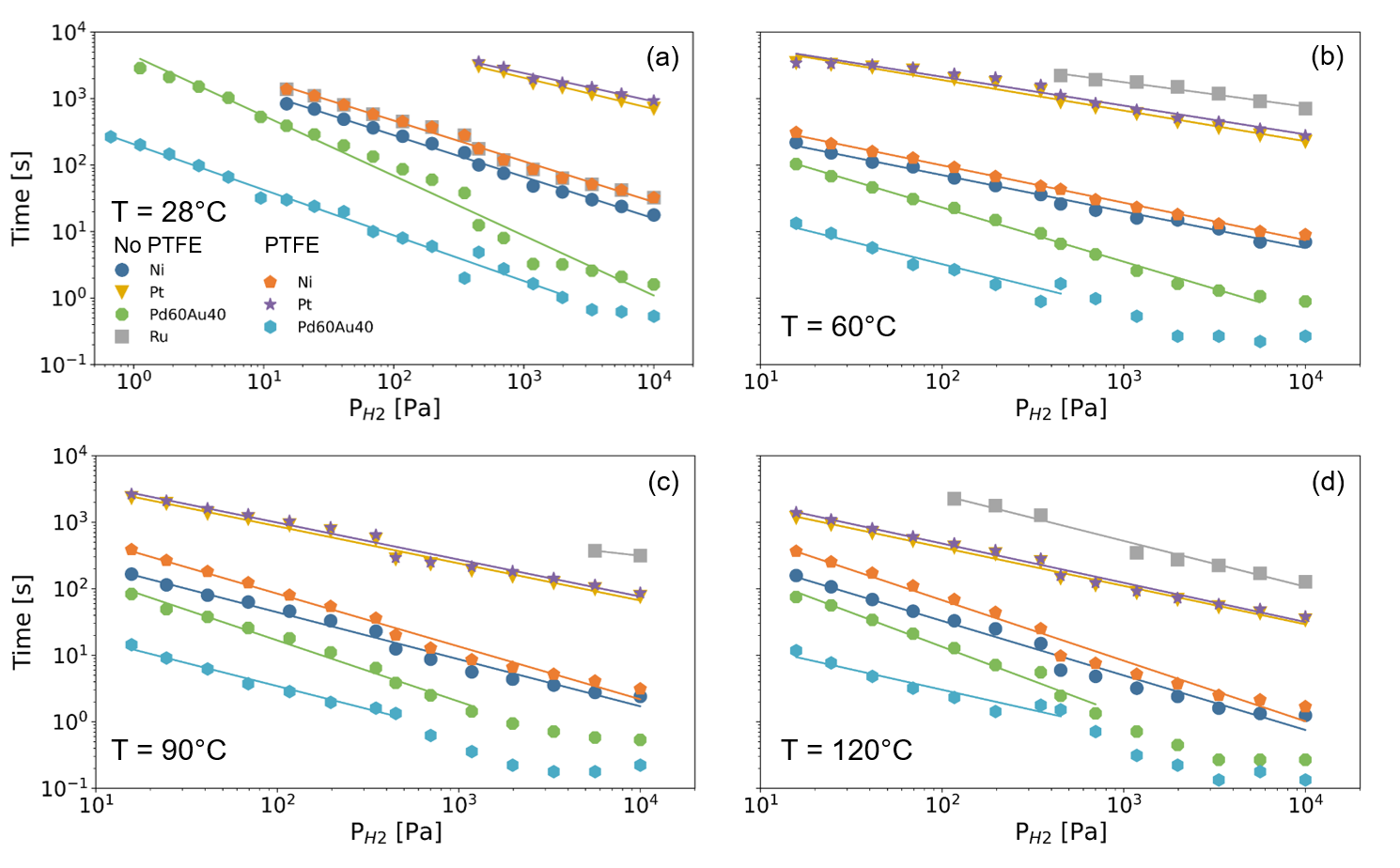}
\caption{Hydrogen pressure dependence of the hydrogen absorption response times of 40~nm Ta thin films capped with 10~nm Ni, Pt, Ru and Pd$_{0.6}$Au$_{0.4}$ at (a,e) $T$ = 28, (b,f) 60, (c,g) 90 and (d,h) 120~$\degree$C (a-d) without and (e-h) with a 30~nm PTFE layer on top of the capping layer. The response time is defined as the time to reach 90\% of the total signal in Fig. \ref{RawIncreasingRT}. The solid lines indicate fits to a power law ($R = a\cdot P_{H2}^\gamma$) for which the exponents $\gamma$ are reported in Table \ref{exponentsincreasing}. }
\label{ResponseTimesIncreasingOtherElements}
\end{center}
\end{figure*}

\begin{figure*}[tb]
\begin{center}
\includegraphics[width= 0.9\textwidth]{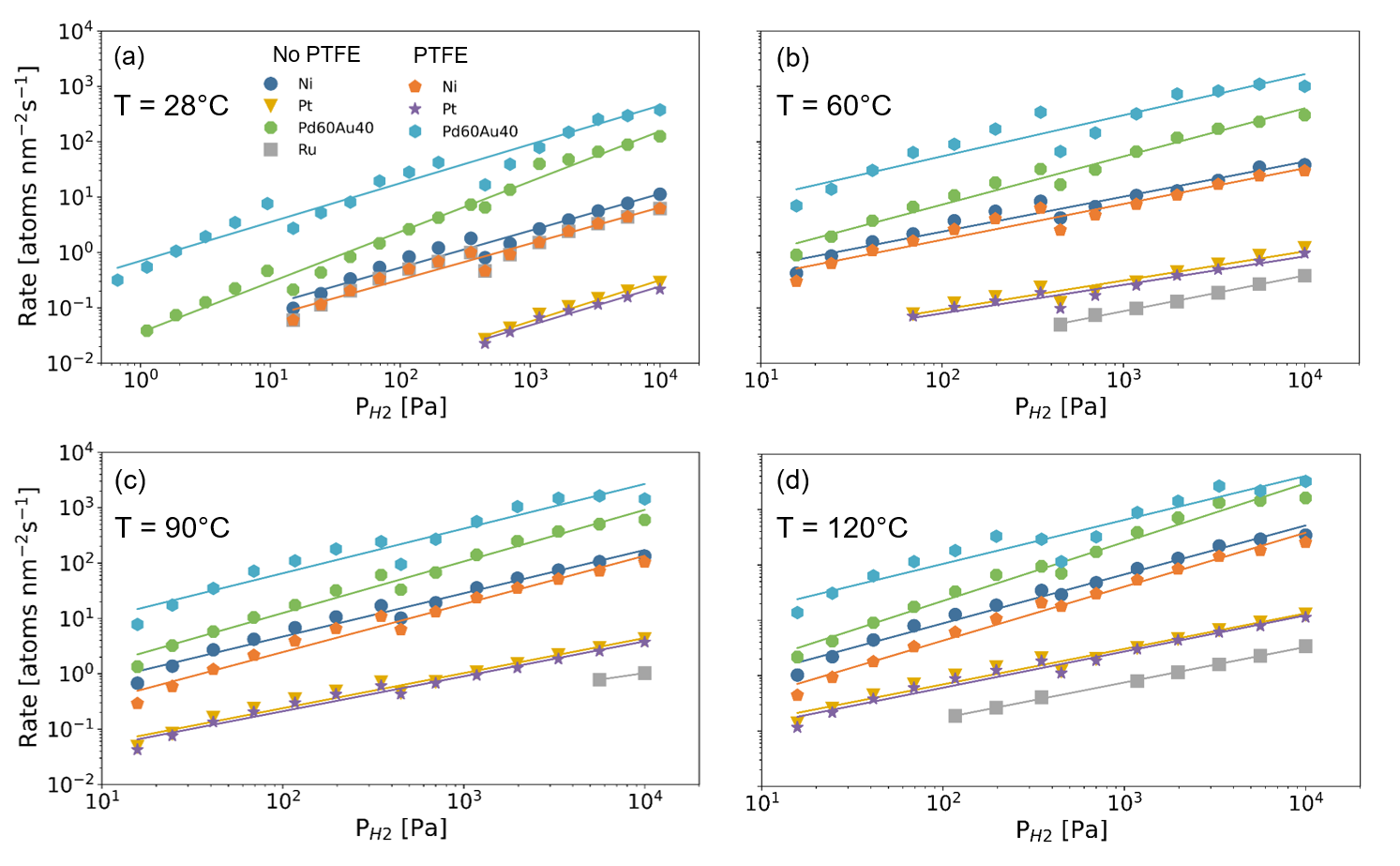}
\caption{Hydrogen pressure dependence of the hydrogen absorption rate of 40~nm Ta thin films capped with 10~nm Ni, Pt, Ru and Pd$_{0.6}$Au$_{0.4}$ at (a,e) $T$ = 28, (b,f) 60, (c,g) 90 and (d,h) 120~$\degree$C (a-d) without and (e-h) with a 30~nm PTFE layer on top of the capping layer. The hydrogen absorption rate $R$ is computed according to eq. \ref{eq1}. The solid lines indicate fits to a power law ($R = a\cdot P_{H2}^\gamma$) for which the exponents $\gamma$ are reported in Table \ref{exponentsincreasing}. }
\label{RatesIncreasingOtherElements}
\end{center}
\end{figure*}

\begin{figure*}[tb]
\begin{center}
\includegraphics[width= 0.85\textwidth]{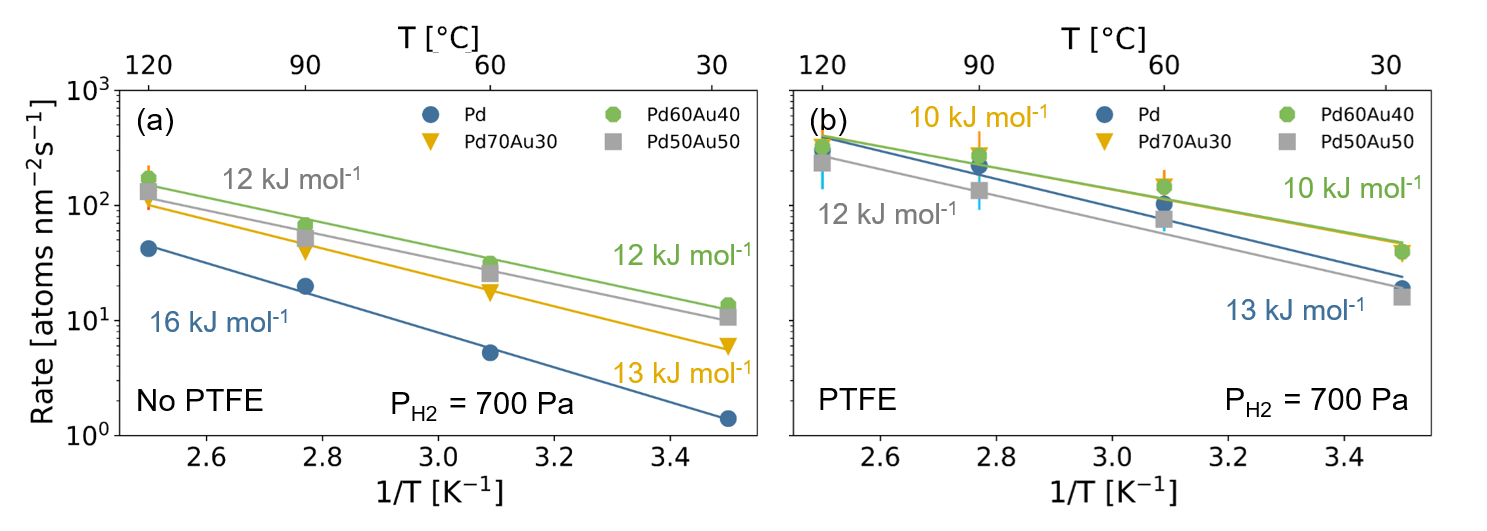}
\caption{Arrhenius plots of the hydrogen absorption rate at $P_{H2}$ = 700~Pa of 40~nm Ta thin films capped with 10~nm Pd$_{1-y}$Au$_{y}$ (a) without and (b) with a 30~nm PTFE layer on top of the capping layer. The solid lines indicate fits to the Arrhenius equation of eq. \ref{arr} and the numbers indicate the apparent activation energy obtained from the fits. }
\label{ArrheniusPd}
\end{center}
\end{figure*}

\begin{figure}[tb]
\begin{center}
\includegraphics[width= 0.45\textwidth]{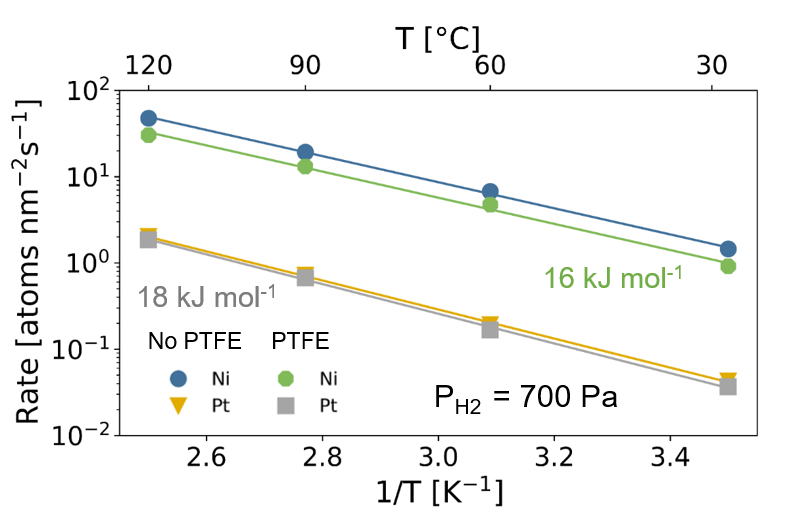}
\caption{Arrhenius plots of the hydrogen absorption rate at $P_{H2}$ = 700~Pa of 40~nm Ta thin films capped with 10~nm Ni and Pt with and without a 30~nm PTFE layer on top of the capping layer. The solid lines indicate fits to the Arrhenius equation of eq. \ref{arr} and the numbers indicate the apparent activation energy obtained from the fits. }
\label{ArrheniusOther}
\end{center}
\end{figure}

\begin{table*}[tb]
\centering
\caption{This table reports fitted exponents $\gamma$ of the fit of the power law $R = a\cdot P_{H2}^\gamma$ to the pressure dependence of the hydrogen absorption rate $R$. The fits are displayed in Figures \ref{PressureDependenceIncreasingRT}(c), \ref{PressureDependenceRateIncreasingHighT} and \ref{RatesIncreasingOtherElements}.}
\label{exponentsincreasing}
\begin{tabular}{llllllllll}
\hline \hline
                     & \multicolumn{4}{c}{No PTFE} &  & \multicolumn{4}{c}{PTFE}  \\ 
										 & 28~$\degree$C   & 60~$\degree$C   & 90~$\degree$C  & 120~$\degree$C &  & 28~$\degree$C  & 60~$\degree$C  & 90~$\degree$C  & 120~$\degree$C \\ \cline{2-5}\cline{7-10}
Pd                   & 0.71  & 0.89  & 1.00 & 1.02 &  & 0.74 & 0.98 & 1.05 & 1.17 \\
Pd$_{0.7}$Au$_{0.3}$ & 0.84  & 0.86  & 0.99 & 1.09 &  & 0.73 & 0.93 & 1.04 & 1.14 \\
Pd$_{0.6}$Au$_{0.4}$ & 0.91  & 0.90  & 1.07 & 1.12 &  & 0.73 & 0.90 & 1.04 & 1.14 \\
Pd$_{0.5}$Au$_{0.5}$ & 0.87  & 0.87  & 1.02 & 1.10 &  & 0.63 & 0.79 & 0.98 & 1.12 \\
                     &       &       &      &      &  &      &      &      &      \\
Ni                   & 0.67  & 0.63  & 0.78 & 0.88 &  & 0.66 & 0.65 & 0.87 & 0.97 \\
Pt                   & 0.75  & 0.53  & 0.63 & 0.64 &  & 0.70 & 0.51 & 0.63 & 0.65 \\
Ru                   & 		   & 0.64  & 		  & 0.64 &  &      &      &      &      \\ \hline \hline
\end{tabular}
\end{table*}

\section{Results}

\subsection{Pd-based capping layers}
\subsubsection{Optical Response and hydrogenation of the capping layer}
We start our empirical analysis by studying the hydrogenation(kinetics) of the Pd$_{1-y}$Au$_y$ capping layers. The rationale to partly substitute Pd by Au is threefold. First of all, a previous study by Namba et al. indicates that alloying the Pd(110) surface with Au may significantly increase the hydrogen absorption kinetics through a reduction of the activation energy \cite{namba2018}. Second, substitution by Au reduces the hydrogenation of the layer at higher hydrogen pressures, and, for sufficiently large Au concentrations ($y$ $\gtrsim$ 0.2), suppresses the first-order metal-to-metal hydride phase transition from the dilute $\alpha$-PdH$_x$ to the higher concentration PdH$_x$ $\beta$-phase, i.e. thus eliminating the associated hysteresis and the sluggish phase transition \cite{luo2010,westerwaal2013,wadell2015,nugroho2018,bannenberg2019PdAu}. This is especially relevant to hydrogen sensing applications as it ensures that the hydrogenation of the capping layer, and thus the (optical) response, is free of any hysteresis. Furthermore, a limited hydrogenation of the capping layer reduces the volumetric expansion, thus enhancing the reduced long-term stability of the layer. In addition, a large hydrogenation of the capping layer itself also increases the total amount of hydrogen to be dissociated at the surface, which may result in larger response times. Third, substitution by Au increases the size of the face-centered cubic unit cell (see Fig. \ref{XRD}(c)), which may together with an increased hydrogen solubility at lower pressures accelerate the transport across the capping layer. Indeed, although Au itself does not absorb hydrogen and the substitution of Pd by Au reduces the hydrogenation at larger hydrogen pressures, the increased unit cell results at lower hydrogen pressures in larger hydrogen solubilities of the capping layer, showing a maximum for $y$ $\approx$ 0.3, and this enhanced solubility may in turn accelerate the hydrogen transport. 

These trends are quantitatively reproduced for the 10~nm capping layers. Figure \ref{OpticalResponse} displays the partial hydrogen pressure dependence of the optical response of our 10~nm Pd$_{1-y}$Au$_y$ capping layers at four different temperatures. Neutron reflectometry measurements \cite{bannenberg2019PdAu} show that the optical response, measured as the natural logarithm of the white light optical transmission $\mathcal{T}$ relative to the transmission of the as-prepared film ($\mathcal{T}_{prep}$), is irrespective of the Au concentration of the alloy directly proportional to the hydrogenation of Pd$_{1-y}$Au$_y$H$_x$ thin films. Most important, no hysteresis (and sluggish phase transitions) can be discerned for the compositions with $y$ $>$ 0.2, which together with the strongly reduced hydrogenation at high hydrogen pressures and the increased hydrogen solubility at low hydrogen pressures make these compositions most suitable as capping layers. On top of that, the reduced hydrogenation of the capping layer and thus the reduced volumetric/thickness expansion as well as the increased adhesion of Pd$_{1-y}$Au$_y$ enhance the long-term structural stability of the film. We note here that the presence of hysteresis and the hydrogenation of the capping layers may depend on the layer thickness and adhesion conditions \cite{feenstra1983,pundt2006,pivak2009,pivak2011,wagner2016,burlaka2016,bannenberg2019PdAu}. 



\subsubsection{Response times}
Next, we investigate the room temperature response times of Pd$_{0.7}$Au$_{0.3}$, Pd$_{0.6}$Au$_{0.4}$ and Pd$_{0.5}$Au$_{0.5}$, i.e. the Pd$_{1-y}$Au$_y$ compositions for which we did not find any signs of hysteresis. As a reference, we also consider widely used Pd. In Fig. \ref{RawIncreasingRT} we display the response of these samples to a series of increasing pressure steps of 150 $<$ $P_{H2}$ $<$ 10,000~Pa without (Fig. \ref{RawIncreasingRT}(a-d)) and with an additional PTFE layer on top of the capping layer (Fig. \ref{RawIncreasingRT}(e-h)). 

The results show three general trends. First of all, we find that the substitution of Pd by Au in the capping layer significantly reduces the response times. The reduction is the largest for Pd$_{0.6}$Au$_{0.4}$ (Fig. \ref{RawIncreasingRT}(c)), showing response times that are about five times shorter than for Pd (Fig. \ref{RawIncreasingRT}(a)). The enhancement of the kinetics is even greater for $P_{H2}$ $>$ 1000~Pa, where the first-order metal-to-metal hydride transition renders the response of Pd highly hysteric and for which response times are in the minutes time scale (not shown). In stark contrast, Pd$_{0.6}$Au$_{0.4}$ has, as a result of the absence of a first-order metal-to-metal hydride transition (Fig. \ref{OpticalResponse}), response times below 2~s for $P_{H2}$ $>$ 1000~Pa. 

Second, we observe that the presence of the PTFE layer on top of the capping layer significantly reduces the response times. This reductions is the strongest for Pd, for which PTFE reduces the response times by a factor of approximately fifteen. With increasing Au concentration, the effect of PTFE on the response times is reduced. However, with a reduction of the response times with a factor of about four for Pd$_{0.5}$Au$_{0.5}$ the effect remains significant. 

Third, we find that the response times are reduced considerably with increasing pressures. Figure \ref{PressureDependenceIncreasingRT}(a,b) displays the pressure dependence of the response time, here defined as the time to reach 90\% of the total signal, at room temperature and for 0.5 $<$ $P_{H2}$ $<$ 10,000~Pa. The variation in response times is considerable: at a pressure of $P_{H2}$ = 1.0~Pa the response time of the Pd$_{0.6}$Au$_{0.4}$ capping layer with PTFE is about 2.0$\times$10$^{2}$~s, while at $P_{H2}$ = 10,000~Pa it has reduced to about 0.3~s. 

To further study the hydrogen absorption kinetics and to identify the rate limiting step in the hydrogen absorption, we consider the hydrogen sorption rate $R$ instead of the response times in Fig. \ref{PressureDependenceIncreasingRT}(c). When the kinetics are purely surface limited and the hydrogen dissociation is thus the rate-limiting step, the rate is proportional to the applied pressure ($R$ $\propto$ $P_{H2}$), whereas if the kinetics are solely dictated by diffusion $R$ $\propto$ $\sqrt{P_{H2}}$ (see, e.g., refs. \cite{uchida1987,schweppe1997,borgschulte2008}). If the rate is limited by a combination of the two, the exponent will be between 0.5 and 1.0.

We compute the reaction rate from the response times $t_{90}$, the thickness of the Ta layer $d$ and the difference in hydrogenation of the TaH$_x$ layer $\Delta x$ obtained from previous neutron reflectometry experiments \cite{bannenberg2019} according to:

\begin{equation}
R = 0.9\frac{d\rho\Delta x}{t_{90}M},
\label{eq1}
\end{equation}   

\noindent where $\rho$ and $M$ are the experimentally determined density and molar mass of the Ta thin film, respectively.

The exponents of the power law $R = a\cdot P_{H2}^\gamma$ that are fitted to the data are reported in Table \ref{exponentsincreasing} and take at room temperature values of $\gamma$ $\approx$ 0.7 for Pd and $\gamma$ $\approx$ 0.9 for Pd$_{0.6}$Au$_{0.4}$ and Pd$_{0.5}$Au$_{0.5}$. From these values, we conclude that especially for Pd the hydrogen sorption is limited both by surface and diffusion effects, whereas especially for the Pd$_{0.6}$Au$_{0.4}$ and Pd$_{0.5}$Au$_{0.5}$ capped samples the rate is limited predominately by surface effects. As such, the results indicate the diffusion of hydrogen through Pd$_{1-y}$Au$_y$ can, possibly as a result of the increased hydrogen solubility at lower pressures and the increased unit cell, be significantly enhanced as compared with Pd. 

The kinetics fasten at increasing temperatures. Comparing data at $T$ = 60~$\degree$C to that at $T$ = 28~$\degree$C, we find for the best performing sample, Pd$_{0.6}$Au$_{0.4}$ with PTFE, and at a pressure $P_{H2}$ = 250~Pa, a reduction from $t_{90}$ = 6.9 to 2.4~s. This is reflected in the hydrogen sorption rates, for which we find an increase of about a factor two per 30~$\degree$C for all samples. Also at elevated temperatures, the pressure dependence for all samples can be satisfactory described by the power law $R = a\cdot P_{H2}^\gamma$. For all elevated temperatures of $T$ $>$ 60~$\degree$C and all Pd$_{1-y}$Au$_{y}$ compositions studied, the exponents take values of $\gamma$ $\approx$ 1. As such, it indicates that the hydrogen uptake is limited by the surface reaction, suggesting that diffusion is enhanced considerably at elevated temperatures and is not limiting the sorption rates. 

To extract the height of the activation barriers from the temperature dependence of the hydrogenation rates, we display in Fig. \ref{ArrheniusPd} the so-called Arrhenius plots of the absorption rates at a constant hydrogen pressure of $P_{H2}$ = 700~Pa for the Pd$_{1-y}$Au$_{y}$ capping layers. The temperature dependence of the rates are well described by the Arrhenius function:

\begin{equation}
R = R_0e^{-E_A/(k_BT)},
\label{arr}
\end{equation}

\noindent where $E_A$ is the apparent activation energy, $R_0$ a constant and $k_B$ Boltzmann's constant. For bare Pd we find an apparent activation energy of $E_A$ = 16~kJmol$^{-1}$, of similar magnitude as the value reported for Pd nanoparticles deposited on a glass support of 20 kJmol$^{-1}$ \cite{stolas2020}.

The activation energies show two important trends. First of all, consistent with previous research \cite{namba2018,nugroho2019}, the activation energy decreases with increasing Au concentration. Compared with Pd, the activation energy is reduced by about 25\% to 12~kJmol$^{-1}$ for Pd$_{0.6}$Au$_{0.4}$ and Pd$_{0.5}$Au$_{0.5}$. Second, in accordance with the observation that PTFE can significantly shorten the response times, the activation energy of all layers with PTFE is reduced by about 2~kJmol$^{-1}$. Consistent with our previous measurements, the best performing capping layer Pd$_{0.6}$Au$_{0.4}$ with PTFE has the lowest activation energy of only 10~kJmol$^{-1}$.


\subsection{Non Pd-based capping layers}
Next, we consider the capping layers that are based on elements different from Pd, i.e. Ni, Pt and Ru, all materials that hardly hydrogenate under moderate conditions. Exemplary response time measurements at room temperature for Ni and Pt with and without a PTFE capping layer are displayed in Fig. \ref{RawIncreasingRT_Ni_Pt}, the corresponding response times in Fig. \ref{ResponseTimesIncreasingOtherElements}, and the computed reaction rates in Fig. \ref{RatesIncreasingOtherElements}. 

Based on these figures we draw two important conclusions. First, we find that the response times of Ni, Pt, and Ru are much longer than for all Pd-based materials. While the room temperature response times of Ni are about 50-300\% longer than Pd, the response times of Pt are about a factor twenty longer (largely depends on pressure), and no complete response was found at room temperature for Ru within one hour. These trends persist at elevated temperatures. For example, at $T$ = 60~$\degree$C and for a pressure of $P_{H2}$ = 700~Pa, the response times are 30~s, 8$\times$10$^2$~s, 2$\times$10$^3$~s for Ni, Pt and Ru, respectively, while the same response is achieved within 5~s for the fastest material studied, Pd$_{0.6}$Au$_{0.4}$. 

The longer response times of Ni, Pt and Ru as compared with the Pd-based materials may stem from slower dissociation of hydrogen at the surface and/or from slower diffusion of hydrogen through the capping layer. To study this, we consider the pressure and temperature dependence of the hydrogen absorption rates. The pressure dependence of the rates can, as for the Pd-based materials, be described by the power law $R = a\cdot P_{H2}^\gamma$. The corresponding fits to the data are indicated in Fig. \ref{RatesIncreasingOtherElements} and the exponents $\gamma$ are tabulated in Table \ref{exponentsincreasing}. The most striking observation is that the exponents are, regardless of temperature, much lower than for the Pd-based materials. Indeed, we find exponents at $T$ = 60~$\degree$C of about $\gamma$ $\approx$ 0.5-0.6, much lower than $\gamma$ = 0.9 for Pd and Pd$_{1-y}$Au$_y$ alloys. As the exponents for Ni, Pt and Ru are relatively low and reasonably close to 0.5, this indicates that the kinetics are largely limited by the diffusion of hydrogen through the capping layer. As such, the long response times of these materials at room temperature may not come as a complete surprise: The diffusion constant of hydrogen in Ru of 2 10$^{-19}$ m$^2$s$^{-1}$ is at room temperature about four orders of magnitude smaller than in Pd thin films\cite{li1996,soroka2020}. While diffusion data on thin films of Ni is to the best of our knowledge unknown, the values reported for bulk suggest that they are of similar order of magnitude at elevated temperature, but the diffusion is at room temperature in Ni substantially slower than in Pd \cite{devanathan1962,ebisuzaki1968,holleck1970,louthan1975}.

In Fig. \ref{ArrheniusOther} we display the Arrhenius plots of the absorption rates at a constant hydrogen pressure of $P_{H2}$ = 700~Pa for Ni and Pt. The temperature dependence of the rates are well described by the Arrhenius function (eq. \ref{arr}), from which we obtain values of the apparent activation energy of $E_A$ = 16 and 18~kJmol$^{-1}$ for Ni and Pt, respectively. Thus, the activation energy of Ni is comparable to the activation energy of Pd while the one of Pt is slightly larger. As such, it indicates that the slower kinetics in Pt can thus partly be explained by the larger activation energy of the hydrogen dissociation process.

Second, Figs. \ref{RawIncreasingRT_Ni_Pt} - \ref{RatesIncreasingOtherElements} show that unlike for the Pd-based capping materials, the PTFE layer hardly affects the response times. Consistently, we find that the activation energies of Ni and Pt are not affected by the presence of PTFE (Fig. \ref{ArrheniusOther}). In the literature, the hydrogenation kinetics enhancing effects of PTFE have been suggested to be due to (i) a lower activation energy of hydrogen dissociation, possibly as a result of strain imposed by the PTFE layer on the capping layer, or (ii) to more active sites on the surface remaining available to dissociate the hydrogen \cite{ngene2014,delmelle2016,nugroho2019}. The results presented in this paper are consistent with the first explanation, as we only see a fastening of the kinetics for those samples for which the activation is reduced.

\section{Conclusion}
In summary, we systematically studied the hydrogenation kinetics of catalyzing capping layer made of several alloys of Pd and Au as well as single-element Pt, Ni and Ru by exploiting the profound changes in optical transmission of capped Ta thin films. Our results demonstrate that doping Pd with Au results in significantly faster hydrogenation kinetics, with response times up to five times shorter than Pd through enhanced diffusion and a reduction of the activation energy. Detailed analysis of the pressure and temperature dependence of the reaction rates reveals that the hydrogen dissociation at the surface is the rate-limiting step for the Pd-based materials. The kinetics of non-Pd based materials turns out to be significantly slower and mainly limited by the diffusion through the capping layer itself. Surprisingly, PTFE was only found to improve the kinetics of Pd-based capping materials and had no significant effect on the kinetics of Pt, Ni and Ru. As such, these results aid in rationally choosing a suitable capping material for the application of metal hydrides in a green economy. In a wider perspective, the method developed in this paper can be used to simultaneously study the hydrogenation kinetics and determine diffusion constants in thin film materials for a wide set of experimental conditions.

\section*{Acknowledgments}
We thank Bernard Dam for fruitful discussions and comments.

\bibliography{Hydrogen_Sensing}

\newpage
\begin{center}
\section*{Supplementary Figures}
\end{center}
\renewcommand{\thefigure}{S\arabic{figure}}
\setcounter{figure}{0}

\begin{figure}[tb]
\begin{center}
\includegraphics[width= 0.45\textwidth]{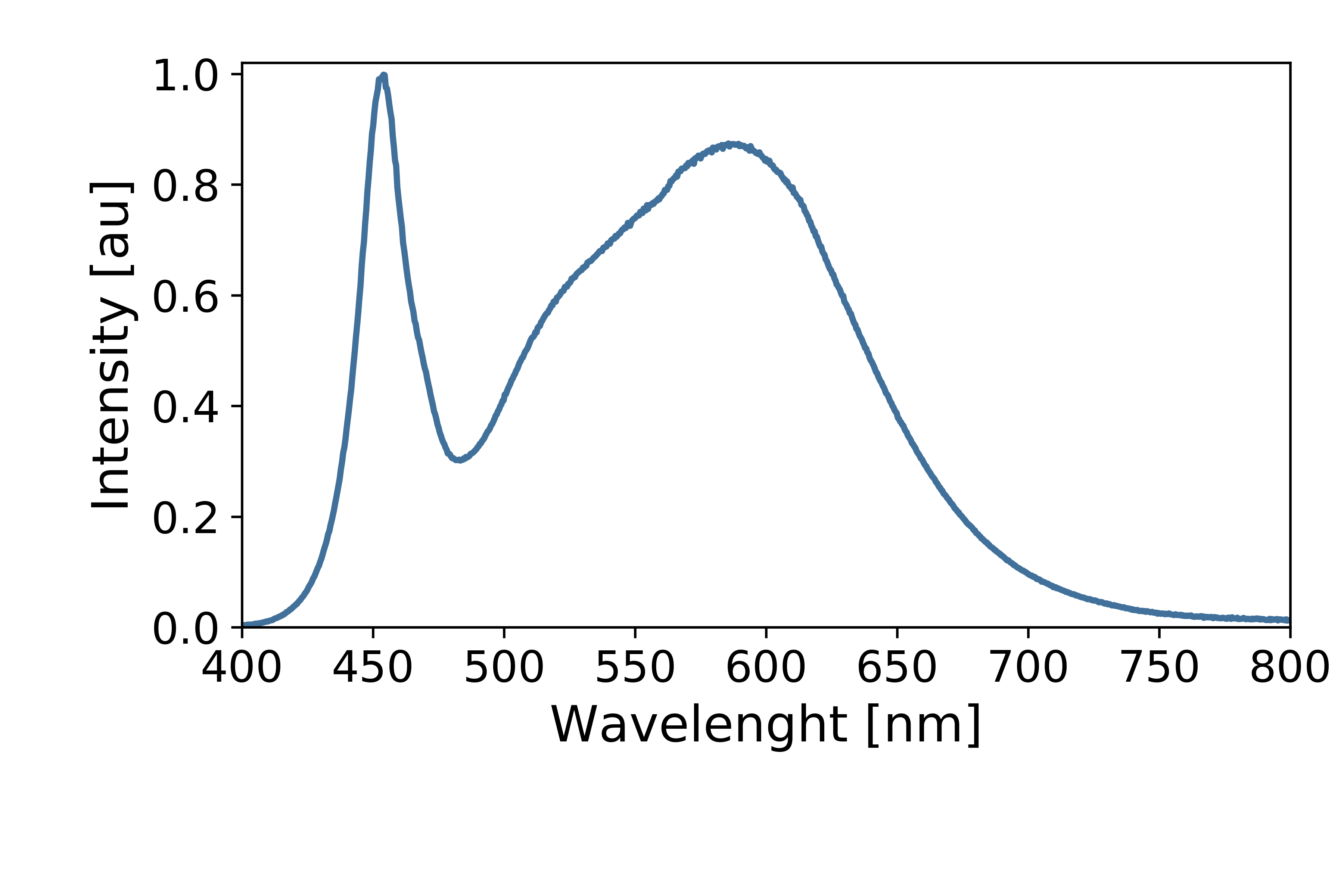}
\caption{Spectrum of the Philips MR16 MASTER LEDs (10/50~W) with a color temperature of 4,000~K used for the white-light optical transmission (hydrogenography) measurements.}
\label{SpectrumLED}
\end{center}
\end{figure}

\begin{figure*}[tb]
\begin{center}
\includegraphics[width= 0.8\textwidth]{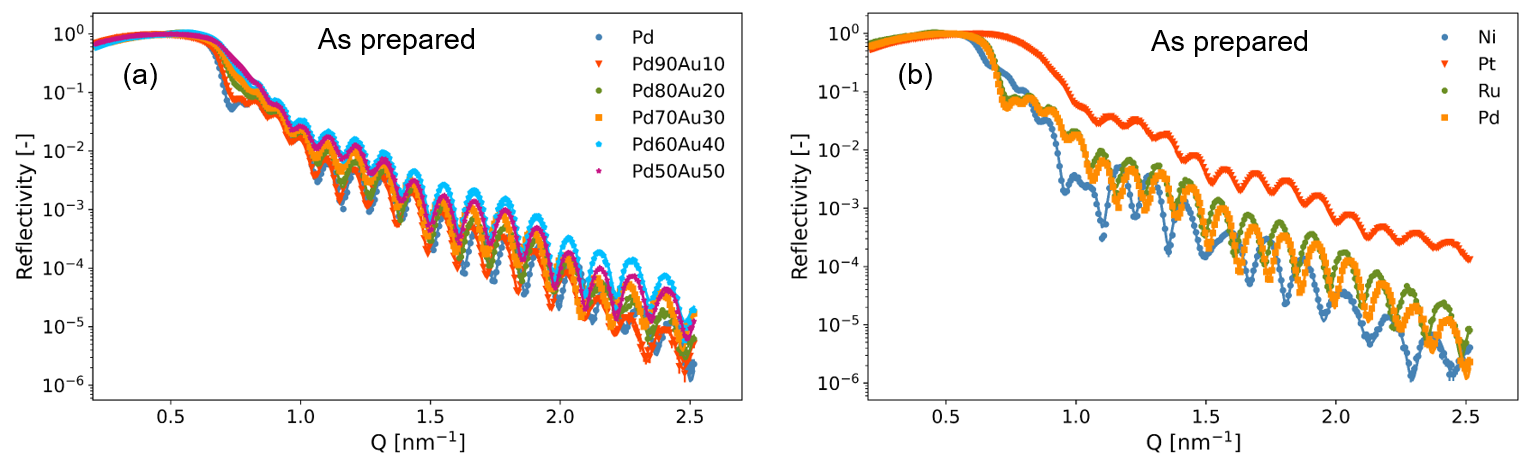}
\caption{X-Ray Reflectometry (XRR) measurements of the as-prepared Ta thin films with a Ti adhesion layer and for the compositions of the capping layers indicated in the figure legend. The solid lines indicate fits to the measured data from which the layer thicknesses, density and roughness of the various layers as reported in Table \ref{XRRfits} have been obtained. }
\label{XRR}
\end{center}
\end{figure*}

\begin{table*}[tb]
\caption{Fitted layer thickness, density and roughness $\sigma$ of the as-prepared Ta thin films with a Ti adhesion layer capped with the various compositions indicated. The fits are reported in Fig. \ref{XRR}. The samples with and without a PTFE layer were produced during the same deposition and have identical composition and layer thicknesses. The density of the fused quartz substrate was fixed to the literature value of 26.5 FU/nm${^3}$ and the roughness was fitted to $\sigma$ = 0.3 $\pm$ 0.1~nm for all samples.}
\label{XRRfits}
\centering
\begin{tabular}{lllllllllllll}
\hline
                  &   & \multicolumn{6}{c}{Pd$_{1-y}$Au$_y$} & & \multicolumn{1}{c}{Ni}   & \multicolumn{1}{c}{Pt}   & \multicolumn{1}{c}{Ru}    \\ 
                  &   & $y$ = 0   & $y$ = 0.1 & $y$ = 0.2 & $y$ = 0.3 & $y$ = 0.4 & $y$ = 0.5 & & &   &     \\ \cline{3-8} \cline{10-12}
									& {Thickness [nm]}         & 10.0 & 10.4     & 10.5     & 10.9     & 11.0     & 9.6     & & 10.9 & 9.6  & 10.4  \\
Cap Layer         & {Density [FU/nm${^3}$]}  & 67   & 67       & 66       & 64       & 62       & 67       & & 91   & 66   & 73.6   \\
                  & {$\sigma$ [nm]}      & 1.0  & 1.1      & 1.0      & 0.9      & 0.8      & 0.9      & &1.3  & 0.6  & 0.9   \\ \hline
									& {Thickness [nm]}         & 36.7 & 36.9     & 36.9     & 37.0     & 36.4     & 37.5    & & 37.2 & 37.5 & 36.7 \\
Ta Layer          & {Density [FU/nm${^3}$]} & 55   & 55       & 55       & 55       & 55       & 55       & & 55   & 55   & 55     \\
                  & {$\sigma$ [nm]}      & 0.9  & 0.9      & 0.8      & 0.3      & 0.2      & 0.5      & & 1.0  & 0.8  & 0.9   \\ \hline
									& {Thickness [nm]}         & 3.8  & 4.2      & 3.9      & 4.2      & 3.8      & 3.9      & &4.3  & 4.2  & 3.9   \\
Ti Layer          & {Density [FU/nm${^3}$]} & 43   & 49       & 51       & 50       & 49       & 55  &     & 47   & 53   & 51     \\
                  & {$\sigma$ [nm]}   & 0.7  & 0.7      & 0.7      & 0.7      & 0.7      & 0.6      & & 0.8  & 0.7  & 0.6   \\ \hline
\end{tabular}
\end{table*}

\begin{figure*}[tb]
\includegraphics[width= 1\textwidth]{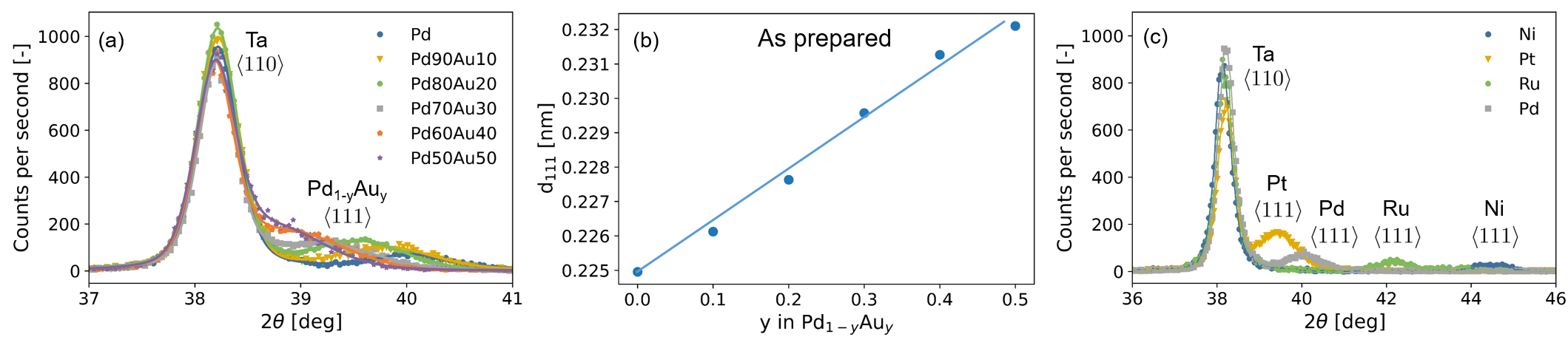}
\begin{center}
\caption{Ex-situ XRD results of the 40~nm Ta thin films with a 4~nm Ti adhesion layer and capped with a 10~nm capping layer. (a) Pd$_{1-y}$Au$_y$ and (b) with Pd layer after exposure of the thin films to hydrogen and measured in air. Diffractograms (Cu-K$\alpha$ $\lambda$ = 0.1542~nm) of the thin films capped with a Pd$_{1-y}$Au$_y$ layer and (b) with a Ni, Pt, Ru and Pd capping layer. The continuous lines represent fits of two pseudo-Voigt functions to the experimental data. (c) Au doping dependence of the d$_{111}$-spacing in Pd$_{1-y}$Au$_y$. The solid line indicates Vergard's law, i.e. $d_{Pd_{1-y}Au_y} = (1-y)d_{Pd} + y d_{Au}$, with the endpoints, $d_{Pd}, d_{Au}$, measured on sputtered thin film samples.}
\label{XRD}
\end{center}
\end{figure*}

\end{document}